\documentclass[aps,prx,onecolumn,showpacs,superscriptaddress]{revtex4-1}
\usepackage{amssymb}
\usepackage{amsfonts}
\usepackage{amsmath}
\usepackage{bm}
\usepackage{textcomp}
\usepackage{color}
\usepackage{longtable}
\usepackage{graphicx}
\usepackage{dcolumn}
\usepackage[a4paper=true,pagebackref=false]{hyperref}
\usepackage[normalem]{ulem}
\usepackage{lineno}

\begin{document}


\title{Improved-Sensitivity Integral SQUID Magnetometry of (Ga,Mn)N Thin Films in Proximity to Mg-doped GaN}

\author{Katarzyna~Gas} \email{kgas@ifpan.edu.pl}
\affiliation{Institute of Physics, Polish Academy of Sciences, Aleja Lotnikow 32/46, PL-02668 Warsaw, Poland}

\author{Gerd~Kunert}
\affiliation{Institute of Solid State Physics, University of Bremen, Otto-Hahn-Allee 1, 28359 Bremen, Germany}

\author{Piotr~Dluzewski}
\affiliation{Institute of Physics, Polish Academy of Sciences, Aleja Lotnikow 32/46, PL-02668 Warsaw, Poland}

\author{Rafal~Jakiela}
\affiliation{Institute of Physics, Polish Academy of Sciences, Aleja Lotnikow 32/46, PL-02668 Warsaw, Poland}

\author{Detlef~Hommel}
\affiliation{Institute of Solid State Physics, University of Bremen, Otto-Hahn-Allee 1, 28359 Bremen, Germany}
\affiliation{Institute of Experimental Physics, University of Wroc{\l}aw, M. Borna 9, Wroc{\l}aw, Poland}
\affiliation{{\L}ukasiewicz Research Network - PORT Polish Center for Technology Development, Stab{\l}owicka 147, Wroc{\l}aw, Poland}

\author{Maciej~Sawicki}
\affiliation{Institute of Physics, Polish Academy of Sciences, Aleja Lotnikow 32/46, PL-02668 Warsaw, Poland}

\date{\today}

\begin{abstract}

Nominally 45~nm~GaN:Mg~/~5~nm~(Ga,Mn)N~/~45~nm~GaN:Mg trilayers structures prepared by molecular beam epitaxy on GaN-buffered Al$_2$O$_3$ substrates are investigated to verify whether the indirect co-doping  by holes from the cladding layers can alter the  spin-spin interaction in (Ga,Mn)N.
The four investigated structures, differing with the Mg doping level, are carefully characterized at a nanoscale by high-resolution transition electron microscopy (HRTEM), energy-dispersive x-ray spectroscopy, and secondary ion mass spectrometry.
Importantly, HRTEM decisively excluded a presence of foreign Mn-rich phases.
Magnetic studies of these structures are aided by the employment of a dedicated experimental approach of the \emph{in situ} compensation of the magnetic contribution from the substrate, allowing an up to about fifty-fold reduction  of this contribution.
This technique, dedicated to these structures,  simultaneously provides a tenfold reduction of long-term instabilities of the output of the magnetometer and lowers the experimental jitter to merely $5 \times 10^{-7}$~emu at 70~kOe, vastly increasing the precision and the credibility of the results of the standard integral (volume) superconducting quantum interference device magnetometry, particularly in high magnetic fields.
The magnetic characteristics of the trilayer structures established here prove identical with the already known properties of the thick (Ga,Mn)N single layers, namely (i) the low temperature ferromagnetism among Mn$^{3+}$ ions driven by  superexchange and (ii) purely paramagnetic response at higher temperatures.
The possible cause of the lack of any effects brought about by the adjacent Mg-doping is a presence of residual Mn in the cladding layers, resulting in the deactivation of the p-type doping intended there.
This finding points out that a more intensive technological effort has to be exerted to promote the co-doping-driven carrier-mediated ferromagnetic coupling in Mn-enriched GaN, especially at elevated temperatures.

\vspace{2cm}

\end{abstract}

\maketitle

\newpage

\section{Introduction}

\baselineskip 15pt

Since the family of nitrides has reached the status of the second most important semiconducting class of materials after Si, the prospect of spintronics functionalization of GaN has gained in significance.
However, the most obvious way of magnetic functionalization of GaN, following the influential concept of Dietl and co-workers \cite{Dietl:2000_S}, by controlled incorporation of transition metals (mainly Mn), proved fruitless despite intensive efforts \cite{Dietl:2015_RMP}.
The prerequisite condition for the room temperature carrier mediated ferromagnetism is a \emph{simultaneous} presence of a large concentration of \emph{randomly} distributed substitutional Mn cations (at least 5\%, or $2\times 10^{21}$~cm$^{-3}$) and a comparably high concentration of mobile holes ($3\times 10^{20}$~cm$^{-3}$). 
Both requirements constitute a certain technological challenge on their own, combined have proved impossible to meet till now.
This difficulty in achieving Mn concentration in such a high rate is rather a low thermodynamic solid solubility limit of Mn in GaN ($2\times10^{17}$~cm$^{-3}$ \cite{Jakiela:2019_JALCOM}).
Instead of forming a random ternary dilute magnetic semiconductor (DMS) compound [(Ga,Mn)N or Ga$_{1-x}$Mn$_x$N, where $x$ is the Mn concentration], Mn atoms tend to agglomerate during the growth into various Mn-rich nano-precipitates.
They exhibit a wide spectrum of coupling temperatures and spin configurations \cite{Zajac:2003_JAP} and are nested in a strongly diluted paramagnetic GaN:Mn host.
This fact calls for an extensive and thorough nano-characterization effort, as the formed Mn-rich nanocrystals can dominate the magnetic properties, particularly at high temperatures.
Although the recently elaborated bulk-oriented synthesis methods allow to surpass the solubility limit by two orders of magnitude \cite{Garcia:1985_JMMM,Pop:1994_MCP,Zajac:2018_PCGCM}, the obtained GaN:Mn crystals remain still another two orders of magnitude short to satisfy the mean field model requirements \cite{Dietl:2000_S}.

Therefore, the highest hopes to obtain a device viable material have been directed to non-thermal equilibrium growth methods,  where a substantial simultaneous Mn and p-type doping seem achievable.
Among those methods, molecular beam epitaxy (MBE) seems to be the most promising \cite{Kunert:2012_APL}.
However, a direct co-doping of GaN with Mn and Mg  at the same time (Mg is the p-type dopant for GaN \cite{Nakamura:2000,Feduniewicz:2005_JCG}) led to the formation of Mn-Mg$_k$, $k=1,2,3$, complexes \cite{Devillers:2012_SR}, or  yielded unsettled results \cite{Jeong:2004_ASS,Dietl:2015_RMP}.
To circumvent these obstacles, following a very successful record of enhancing the carrier mediated ferromagnetism in other DMS systems \cite{Haury:1997_PRL,Boukari:2002_PRL,Wojtowicz:2003_APL83}, a range of structure designs were attempted to physically separate the Mn- and p-type doping.
In the first class of attempts the effort was aimed at achieving an efficient carrier transfer across the (Ga,Mn)N/p-GaN interface to induce the ferromagnetic (FM) ordering in the former \cite{Arkun:2004_APL}.
In the other one, the hole wave functions were expected to extend into (Ga,Mn)N to facilitate the FM coupling \cite{Nepal:2009_MRS-P}, extending up to 8-9~nm, i.e. well beyond the interfacial region \cite{Tropf:2016_JCG}.
Yet another approach relied on a geometrical confinement, so that a forced coexistence of both, the Mn species and the holes in (Ga,Mn)N based nanorods or nanowires was expected \cite{Lin:2017_NRL}.
Again, also all of these approaches yielded rather conflicting results.
However, it should be noted that in the latter case, a substantial strain specific to the nanowire geometry can substantially increase the Curie temperature of the embedded Mn-rich nanocrystals \cite{Kaleta:2019_NL}.

Due to the absence of any following extensive works reporting either an operational spintronic device or a magnetic phase diagram up to room temperature, it is reasonable to assume that high-temperature FM in GaN with Mn is not under control.
Therefore, it remains likely that the observation of the reported high-temperature magnetic features signalizes an inhomogeneous distribution of Mn atoms formed by either coherent chemical separation or crystallographic phase precipitation.
A very similar conclusion can be drawn from a recent attempt to induce the high temperature FM in GaN by  delta doping ($\delta$-doping) \cite{Wadekar:2020_JALCOM}.
It should be noted, however, that in this case the claims about the above-room-temperature FM are partially based upon non-reproducible results related to unspecified instrumentation issues (c.f. Fig.~1~c of ref.~\cite{Wadekar:2020_JALCOM}).
Finally, it is also worth acknowledging significant  challenges connected with an unambiguous analysis of the results obtained for magnetic layers grown by the  $\delta$-doping method.
A wide ranging and a truly in-depth structural characterization at the nanoscale of Fe  $\delta$-doped GaN was required to unravel the form and chemical composition of Fe-rich nanocrystals coherent with the GaN lattice, whose weak magnetic signatures have been observed even beyond room temperature \cite{Navarro:2019_Crystals,Navarro-Quezada:2020_Materials}.

On the other hand, in single-crystalline (Ga,Mn)N epitaxial films with an experimentally documented random distribution of Mn atoms and a small concentration of compensating defects or impurities, clear FM features were found exclusively at low temperatures \cite{Sarigiannidou:2006_PRB,Sawicki:2012_PRB,Kunert:2012_APL}.
In the low temperature regime, the FM obeys a clear phase diagram and is so robust that the critical exponents for the paramagnetic-ferromagnetic phase transitions were experimentally determined \cite{Stefanowicz:2013_PRB}.
These layers are devoid of carriers \cite{Yamamoto:2013_JJAP,Kalbarczyk:2019_JALCOM} and the Fermi level is pinned on the only partially filled Mn$^{3+}$ level in the mid-gap region of GaN \cite{Wolos:2004_PRB_69, PiskorskaH:2015_JAP,Janicki:2017_SR}.
In the absence of carriers, the necessary Mn-Mn coupling is provided by superexchange.
This major coupling mechanism in wide gap dilute magnetic semiconductors, which is known to be predominantly antiferromagnetic there, turns ferromagnetic for Mn$^{3+}$ ions in tetrahedral configuration \cite{Blinowski:1996_PRB}, i.e. in GaN.
It is the short-range nature of this coupling mechanism which limits the Curie temperature $T_{\mathrm{C}}$ to a mere 10~K for currently maximum achievable $x$ of 10\%. Yet, the industrial importance of the nitride compounds, their current dominance in optoelectronic \cite{Morkoc:2009} and high-power applications \cite{Amano:2018_JPDAP} still holds the doors wide open for a technology-viable nitride semiconductor exhibiting robust magnetic properties above the room temperature.

In this study, we tested the hypothesis of whether indirect p-type co-doping can stimulate an increase in FM ordering temperature in single-phase (Ga,Mn)N.
To this end a very thin (Ga,Mn)N layer is sandwiched between two thicker GaN:Mg cladding layers to form a trilayer heterostructure.
Since the (Ga,Mn)N layer is grown following our elaborated specifications, which routinely results in single-crystalline films of single-Kelvin coupling temperature \cite{Kunert:2012_APL,Stefanowicz:2013_PRB}, any impact of the neighboring holes should yield a straightforward result.
While keeping nominally the same growth parameters for the (Ga,Mn)N central layer, the magnitude of the Mg doping level in the cladding GaN:Mg is substantially varied.
Yet it is kept sufficiently low not to induce polarity inversion and deterioration of the interface morphology \cite{Tropf:2016_JCG}.
The combined thorough structural characterization confirms crystallographically and magnetically single phase of (Ga,Mn)N and the desired layout of our heterostructures.
A dedicated method of the \emph{in situ} compensation of the signal from the substrate has been elaborated and actively employed for the magnetic measurements in our magnetometer \cite{Gas:2019_MST}.
The resulting, up to 50 fold, reduction of the contribution of the substrate allows to sizably increase the real resolution and the credibility of the results.
Basing on these measurements any presence of the "FM-like" signal at elevated temperatures has been excluded -- the layers are found to exhibit purely paramagnetic response at $T \geq  50$~K.
Conversely, clear FM features are found only below about 6~K with the main attributes identical to those already known of the thick (Ga,Mn)N single layers.
Our experiments clearly confirm that the presence of the neighboring GaN:Mg layers did not affect the spin-spin interactions in the central (Ga,Mn)N layer.
The nano-characterization effort, which is unprecedented for such structures, indicates the presence of residual Mn in the GaN:Mg cladding layers.
It is argued that this astray Mn is the most likely culprit responsible for the lack of Mg-doping effect.
This findings indicates that a more intensive technological effort is needed to break the deadlock in the effectiveness of the simultaneous Mn- and Mg-codoping in GaN, when an increase of the strength of the spin-spin coupling in (Ga,Mn)N is targeted.

\section{Experimental}

\subsection{Sample growth}

The nominally (45 nm)~/~(5 nm)~/~(45 nm) trilayer heterostructures GaN:Mg~/~(Ga,Mn)N~/~GaN:Mg investigated here are grown by MBE technique in an EPI 930 MBE chamber on fully relaxed GaN templates (thickness $\sim2 \mu$m) fabricated by MOVPE on $c$-plane 0.3~mm thick Al$_2$O$_3$ substrates.
Radio-frequency nitrogen plasma source operating at a power of 300 W is used and is set to give the nitrogen flux of 1.3~sccm.
Suitable growth parameters for interfaces between GaN:Mg and (Ga,Mn)N, precluding polarity inversion from originally Ga-face to N-face GaN were determined in earlier studies \cite{Tropf:2016_JCG}.
During the growth of both materials metallic films form on the surface.
To create sharp interfaces between the layers of different doping, the metallic films are evaporated by a 700~s growth break at the relevant growth temperatures.
The magnetic Ga$_{1-x}$Mn$_{x}$N middle layers are fabricated with standard growth parameters calibrated to obtain Mn concentration, $x$, of about 5\% on undoped GaN/Al$_2$O$_3$, i.e. at the growth temperature of $730^o$C [10].
The Mg content in the GaN:Mg cladding layers is varied by changing the Mg flux ($\Phi_{\mathrm{Mg}}$) from  0, through $7 \times 10^{-9}$ (low, L) and $1 \times 10^{-8}$ (medium, M) to $2 \times 10^{-8}$ (high, H) Torr beam equivalent pressure (BEP).
All the cladding layers are grown at the same $730^o$C  and the Ga flux of $2 \times 10^{-6}$ Torr BEP.
For the (Ga,Mn)N layers, the Ga flux is $8.5 \times 10^{-7}$ Torr BEP,  with the Mn flux set to $1 \times 10^{-6}$ Torr BEP.
Structures are labeled according to the intensity of $\Phi_{\mathrm{Mg}}$ and listed in Table~\ref{tab:Samples}.

\begingroup

\begin{table}[htb]
  \centering \caption{Data related to the investigated GaN:Mg~/~Ga$_{1-x}$Mn$_{x}$N~/~GaN:Mg trilayers. The sample codes indicate the intensity of the Mg flux used to grow either of the cladding GaN:Mg layers. The columns display the magnitudes of the Mg flux ($\Phi_{\mathrm{Mg}}$), Mn content, ($x$ - obtained from magnetization data), Mg concentration (n$_{\mathrm{Mg}}$) established by secondary ion mass spectrometry, and Curie temperature $T_{\mathrm{C}}$.}
\begin{tabular}{ccccc}
\\
	\hline
    \hline
      Sample &  $\Phi_{\mathrm{Mg}}$   & $x [\%] $  &  n$_{\mathrm{Mg}}$ & $T_{\mathrm{C}}$ [K]\\
      code & $ \,\,\,[10^{-8}$~Torr BEP]\,\,\, & \,\,\,$(d=4.3$~nm)\,\,\, & $ \,\,\,[10^{20}$~cm$^{-3}]\,\,\, $ & \,\,\,$\pm 0.5$~K\,\,\ \\
            \hline
       S$_{\mathrm{0}}$ & $-$ & $4.5 \pm 0.6$ & 0.003 & 2.6 \\
       S$_{\mathrm{L}}$ & 0.7 & $6.4 \pm 0.8$ & 3 & 3.5 \\
       S$_{\mathrm{M}}$ & 1.0 & $4.0 \pm 0.5$ & 4 & 4.1 \\
       S$_{\mathrm{H}}$ & 2.0 & $0.2 \pm 0.2$ & 9 & $<2$ \\
%
	\hline
	\hline
\end{tabular}
  \label{tab:Samples}
\end{table}
\endgroup

\subsection{Structural characterization}


The structural characterization is performed using a FEI Titan 80-300 Cubed Cs image corrected transition electron microscope (TEM) operated at 300 kV and equipped with Energy-dispersive x-ray spectroscopy (EDX) spectrometer.
The cross-sectional lamellas for TEM are prepared by FIB technique with Pt as a protected capping layer.
Element specific depth profiling is performed by secondary ion mass spectrometry (SIMS) using a CAMECA IMS6F microanalyzer.

\begin{figure}[htb]
\centering
\includegraphics[width=15cm]{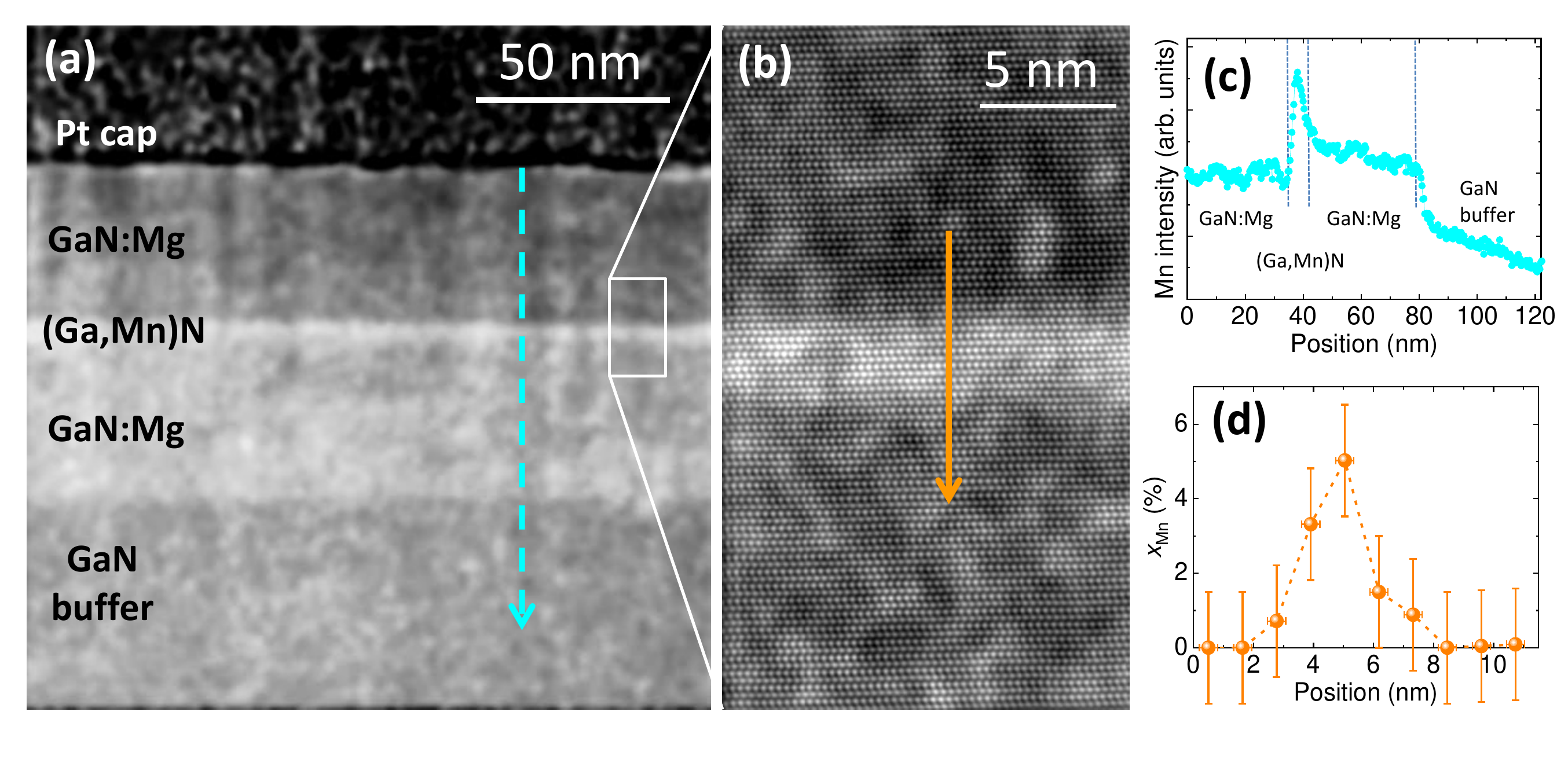}
\caption{\label{Fig:TEM} (Color on line) (a) An example of cross-sectional transmission electron microscopy (TEM) image of the S$_{\mathrm{M}}$ structure (detailed in Table~\ref{tab:Samples}). The Mn-containing layer is running horizontally through the image between GaN:Mg ones. (b) High resolution TEM close up into the (Ga,Mn)N layer and its vicinity. (c) Energy-dispersive x-ray spectroscopy (EDX) established Mn concentration scan along the dashed arrow indicated in (a). (d) EDX-established Mn concentration scan along the solid orange arrow indicated in (b).}
\end{figure}
The accurateness of the obtained heterostructures is confirmed in Fig.~\ref{Fig:TEM}, taking the  S$_{\mathrm{M}}$ structure as the sole representative.
As shown later, this sample exhibits the strongest magnetic response, and so it constitutes the key element to formulate the final conclusions of this study. The TEM image in Fig.~\ref{Fig:TEM}~(a) confirms the trilayer's intended design and a general high quality of the obtained material.
Some incidental defects have been detected, but they are clearly associated with dislocations propagating from the buffer~/~substrate interface to the free surface.
This is a characteristic feature of the heteroexpitaxy of GaN on Al$_2$O$_3$ and so is not presented here.
The HRTEM images, as the one in Fig.~\ref{Fig:TEM}~(b) reveal the high quality crystal structure of (Ga,Mn)N and a high quality of the interfaces.
These images also allow to determine the (Ga,Mn)N layer thickness, $d$.
It is found to range between 3.8 and 4.8~nm [equivalently between 16 and 18 of 00.2 monolayers of  (Ga,Mn)N], and a final value of $d=(4.3 \pm 0.5)$~nm has been adopted for all the structures.
The uncertainty of $d$ given here constitutes the main contribution determining the magnitude of error in the determination of $x$ from magnetic measurements.
According to cross-section EDX measurements, the Mn concentration  in the middle of the layer does not exceed 5\%, Fig.~\ref{Fig:TEM}~(d), but the concentration profile assumes rather a bell-like shape than required rectangular one in the expected location of the (Ga,Mn)N layer.
Given a rather substantial uncertainty of this method, we accept this result as another solid confirmation of the correctness of the growth.
More importantly, basing on the overall HRTEM effort exemplified in Fig.~\ref{Fig:TEM}, we confirm the single-phase character of the investigated multilayers, i.e. the TEM studies excluded the presence of any secondary crystalline phases or Mn aggregations.

\begin{figure}[htb]
\centering
\includegraphics[width=7.cm]{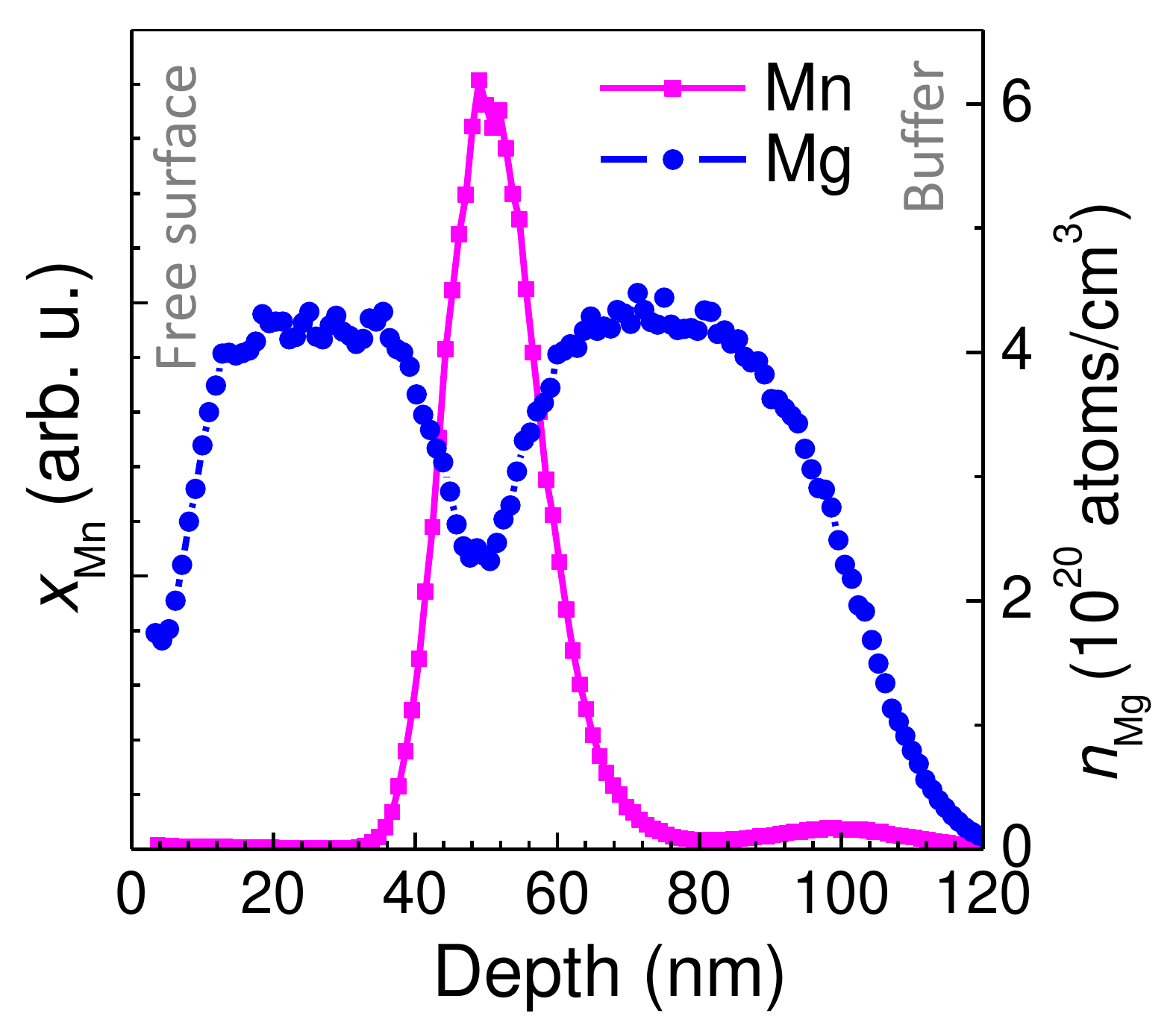}
\caption{\label{Fig:SIMS} (Color on line) Secondary ion mass spectrometry (SIMS) Mn- and Mg-depth profiles of the S$_{\mathrm{M}}$ structure (grown with the medium Mg flux of $1\times 10^{-8}$~Torr BEP). An apparently larger width of the (Ga,Mn)N layer in SIMS than in TEM study is the method's artefact due to mixing/segregation mechanism specific for SIMS measurements \cite{Wilson1989secondary}. }
\end{figure}
The intended Mn and Mg doping profile is further confirmed by SIMS, exemplified in Fig.~\ref{Fig:SIMS} for the same sample S$_{\mathrm{M}}$.
We confirm that all four structures exhibit qualitatively identical profiles. 
SIMS studies confirm that the thin (Ga,Mn)N layer is sandwiched between two, uniformly doped, about 40 nm thick GaN:Mg layers.
The few nm smearing is a typical technique-related artefact originating in mixing and segregation mechanisms \cite{Wilson1989secondary}.
The established Mg doping levels, n$_{\mathrm{Mg}}$, for each of the samples are listed in Table~\ref{tab:Samples}.
It varies form $3\times10^{17}$~cm$^{-3}$ for the sample grown with no Mg flux (S$_{\mathrm{0}}$) to $9\times 10^{20}$~cm$^{-3}$ for the structure grown under high Mg flux (S$_{\mathrm{H}}$).

Importantly, both TEM and SIMS techniques detect a noticeable Mn concentration away from its intended nesting place.
A clear difference in brightness of both GaN:Mg cladding layers are seen in the image in Fig.~\ref{Fig:TEM}~(a).
The vertical 120~nm EDX scan [Fig.~\ref{Fig:TEM}~(c)] through the cross-sectional image in Fig.~\ref{Fig:TEM}~(a), univocally points to an increased Mn concentration in the first-grown cladding layer.
This effect is seen well when comparing with the Mn count from the GaN buffer layer.
The SIMS scan specifically points to the beginning of the growth of the whole trilayer structure, where the increased yield of Mn signal is seen, Fig.~\ref{Fig:SIMS}.

The authors admit that at the moment no feasible explanation can be put forward.
Usually, such or similar effects are assigned to interdiffusion between various constituencies of the structure.
A good example is diffusion of Ta in magnetic tunnel junctions.
It causes the degradation of the perpendicular magnetic anisotropy of the magnetic active layer \cite{Yamanouchi:2011_JAP}.
This detrimental effect can be mitigated by an insertion of a "dusting" sub-nm Mo or W layer to serve as diffusion stopping layers \cite{Kim:2015_SR,Almasi:2016_APL}.
However here, the (out-)diffusion of Mn, as the leading effect, can be excluded.
Firstly, according to the available data \cite{Jakiela:2019_JALCOM} the solubility of Mn in GaN during the diffusion close to the thermal equilibrium at temperatures between 900 and $1100^o$C is 100 to 1000 times smaller than the concentration of Mn found by SIMS at the trilayer/GaN buffer interface.
We evaluate the maximum concentration of the Mn accumulation by diffusion process at about 10~nm slab of the material at the interface region to correspond to only 0.0025\%.
It is expected that at the growth temperature of around $730^o$C it should be at least an order of magnitude lower.
Secondly, the same studies \cite{Jakiela:2019_JALCOM} pointed out that the Mn diffusion through GaN is carried out by the interstitial-substitutional mechanism through Ga vacancies.
However, when the Fermi energy is lowered, and this ought to be the case of GaN:Mg, the formation energy of such defects is very high \cite{Lyons:2017_NPJ-CM,Heikkinen:2020_JCG}.
So the presence of Mg  reduces the number of Ga vacancies in the Mg-doped cladding layers, rendering an appreciable accumulation of Mn at the interface with the GaN buffer via Mn (out-)diffusion from the central (Ga,Mn)N layer even less likely than estimated above.

Another important issue is a possibility of Mg diffusion from the cladding layers to the central layer.
Its importance stems from a large tendency to form Mn-Mg$_k$ complexes in the GaN lattice in which Mn charge- and spin-state are lowered \cite{Devillers:2012_SR}.
These effects starts to be statistically relevant only when the Mg/Mn ratio, $\kappa > 0.5$ \cite{Devillers:2012_SR}.
So, this could be an important effect to reckon with if the total number of Mg atoms in the central (Ga,Mn)N layer exceeded a half  of Mn.
However, this is rather unlikely here.
Firstly, in our structures the maximum concentration of Mg in the cladding layers reaches just a half of the intended Mn concentration.
So $\kappa < 0.5$ even in the worst case scenario (sample  S$_{\mathrm{H}}$), and an existence of Mn-Mg$_k$ complexes cannot be hold responsible for a particularly low $x$ in this sample.
Secondly,  the  diffusion of Mg in GaN is far less vigorous than that of Mn \cite{Benzarti:2008_JCG,Narita:2019_APEx}, so for the growth temperatures relevant here the maximum concentration of the Mg atoms in the central (Ga,Mn)N layer should be lower than $10^{14}$~cm$^{-3}$, and therefore of no importance to the effects considered here.

\subsection{Magnetic measurements and discussion}

It should be pointed out here that for such very thin layers of a highly diluted compound, both EDX  and SIMS techniques can only be regarded as qualitative probes when it comes to the determination of the concentration of the magnetic species.
In order to quantitatively assess the magnitude of $x$ in the Ga$_{1-x}$Mn$_x$N layers, we resort to magnetic measurements, which allow to establish $x$ from the magnitude of the low temperature saturation magnetization.
However, taking into account a rather slowly saturating magnetization, $M$, in (Ga,Mn)N and the documented existence of a sizable magnetic anisotropy, MA \cite{Stefanowicz:2010_PRBa,Sztenkiel:2016_NC,Sztenkiel:2020_NJP}, a rather strong magnetic field of few tens of kOe is required \cite{Gosk:2005_PRB,Stefanowicz:2010_PRBa,Bonanni:2011_PRB,Gas:2018_JALCOM}.
For reasons outlined below, this makes the whole undertaking very challenging.

We start from the notion that for the technological parameters of our (Ga,Mn)N mid-layer the saturation magnetic moment of a $5 \times 5$~mm$^2$ piece at 2~K and 70~kOe should amount to about 8~$\mu$emu.
From one side this value safely exceeds the declared sensitivity of commercial superconducting quantum interference device (SQUID) magnetometers by two orders of magnitude, but from the other, it is only about 1\% of the diamagnetic signal exerted at the same time by the equivalent 0.3~mm thick sapphire substrate, about 1000~$\mu$emu.
Therefore, an exceptional care and adequateness of the experimental procedures have to be enforced to eliminate artifacts and evade limitations associated with the volume SQUID magnetometry \cite{Sawicki:2011_SST,Pereira:2017_JPDAP}.
In the case of this study an efficient elimination of the magnetic contribution brought about by the substrate is the most important factor determining the credibility of the final results and in turn of the whole effort.

Unfortunately, the ubiquitously employed assumption that the magnetic response of the common substrates is "ideally diamagnetic", that is linear in $H$ and $T$-independent, does not correspond to the experimental reality.
To the authors' view, this groundlessly assumed ideal diamagnetism of bulky substrates may be one of the  very significant sources behind the so frequently reported "room temperature ferromagnetism" in GaN doped with transition metals and in similar thin-layer DMS systems.
In this context, the following two leading contributions should be pointed out.

Firstly, the residues of the metallic glue or substrate backside metallization used to attach and/or thermalize the substrates in the growth chambers.
Obviously, the magnitude of this response varies a great deal, but can easily approach 10~$\mu$emu at room temperature, exhibiting an overall sigmoidal or Lagngevin-like shape of its field dependency with a saturation field between 5 and 10~kOe.
An example of such a response can be found in \cite{Gas:2021_arXiv_NWs}.
It should be noted that such a response alone (i) is comparable to the magnitude of magnetic response expected here, and (ii) that it could constitute a solid, however highly not-legitimate basis to claim the existence of an above-room temperature ferromagnetism.
Therefore, this backside material has to be meticulously removed before the measurements.

Secondly, magnetically active contaminants present in the bulk of the substrate.
In the case of sapphire, this surplus "magnetism" is predominantly caused by Cr \cite{Przybylinska:2006_MSEB}, but other transition metals are also likely to be present.
The overall picture is such that the magnetic purity of common substrates does vary from vendor to vendor and it fluctuates in long runs in time, even when the material is acquired from the same source.
Therefore, it is  imperative that only the best, preselected batches of the substrates should be used when magnetic measurements are expected.
However, the best does not mean the ideal.
An extra nontrivial $T$-- and $H$--dependent contribution is always present in the  magnetic moment $m(T,H)$.
The existence of such disturbing FM-like (that is non-linear and hysteretic in $H$) contributions in epi-ready sapphire substrates has been already exemplified in ref.~\onlinecite{Stefanowicz:2010_PRBa} and in the Supplementary Information part of ref.~\cite{Gas:2019_MST}.
Pretending that such a response originates from a hypothetical 5~nm thin Ga$_{1-x}$Mn$_x$N layer, i.e. assuming the ideally diamagnetic response of the sapphire, one can readily end up with $x$ as high as 7 to 10\% on the account of the analysis of the low temperature $m(T,H)$ data obtained for a typical clean sapphire substrate, c.f.~Fig.~S1~(b) of ref.~\cite{Gas:2019_MST}.
Most importantly, basing on the high-$T$ data presented in Fig.~S2~(a) of ref.~\cite{Gas:2019_MST} one could enthusiastically declare an existence of a room temperature FM induced in our trilayer system with the Mn saturation magnetization in the range between $0.6 - 1$~$\mu_{\mathrm{B}}$/Mn.

\begin{figure}[htb]
\centering
\includegraphics[width=16 cm]{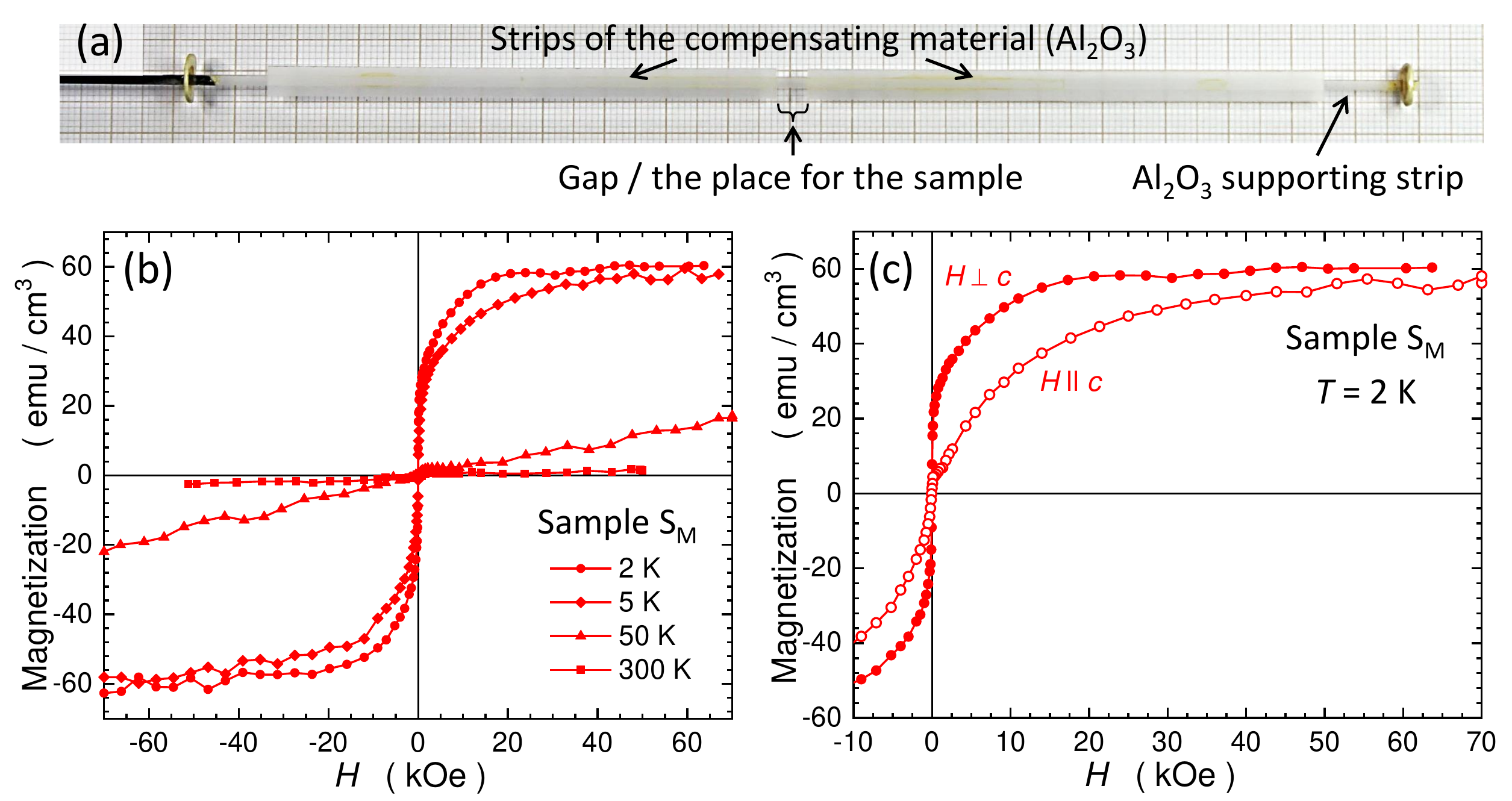}
\caption{\label{Fig:M(H)} (Color on line) (a) The photograph of the compensational sample holder used in this study. The active \emph{in situ} compensation of the substrate of the sample is realized by the side-mounted sapphire strips cut from a 2" sapphire wafer originating from the same batch as the substrates used for the growth of the structures investigated here.  (b) In-plane isothermal magnetization curves $M(H)$ of sample S$_{\mathrm{M}}$ measured at selected temperatures. (c) Magnetic anisotropy at 2 K for the same sample.  The in-plane orientation corresponds to $H$ perpendicular to the wurtzite $c$ axis, $H \perp c$  (bullets), the perpendicular orientation corresponds to $H$ parallel to $c$, $H \parallel c$ (circles), as labeled in the figure. The typical magnitudes of the error bars specific for the $M(H)$ measurements are for clarity of the presentation exemplified in the next figure.}
\end{figure}

In order to mitigate these issues and so to vastly increase the credibility of experimentally established magnetic moments specific to the magnetic layer of interests, $m_{\mathrm{M}}$, a special \emph{in situ} substrate-compensating experimental method was elaborated.
Both, a special (substrate) compensational sample holder (CSH) has been prepared and the adequate experimental routine has been developed \cite{Gas:2019_MST}.
The actual CSH used in this experiment is presented in Fig.~\ref{Fig:M(H)}~(a).
The substrate-compensating strips were cut from a 2" sapphire wafer originating from the same batch of wafers used to deposit the structures studied here.
Accordingly, all the results presented below are substantially devoid of any magnetic contributions not-specific to the (Ga,Mn)N layers of interest.
This method, as a whole undertaking, constitutes a vast experimental advancement.
Firstly, the signal of a particular substrate is compensated down to a single \% level, so  the detrimental "magnetic" contributions of the substrate are reduced in the same proportion, typically between 10 to 20 times.
The full compensation, if needed, can be accomplished upon results of complimentary measurement(s) of a reference piece of the same substrate mounted in the same CSH.
The resulting magnitudes of $m_{\mathrm{M}}(H)$ are established upon the difference of the appropriate scaled results of these two sets of measurements.
The details are given in \cite{Gas:2019_MST}.

Not less importantly the method provides a tenfold reduction of the long-term temporal fluctuations of the magnitude of the magnetometer output.
These instabilities have been found by laboriously probing the same test specimen equivalent to a $5 \times 5 \times 0.3$~mm$^3$ sapphire substrate at nominally identical experimental conditions ($T=300$~K and $H=20$~kOe) in the course of about two years.
As documented in Fig.~2 in ref.~\cite{Gas:2019_MST} the average amplitude of these changes amounts to about 1.5~$\mu$emu, or about 5~$\mu$emu when scaled to the maximum field of 70~kOe.
Such fluctuations of the output level can be seen on a weekly basis (at least the test measurements were not performed much often), the overall two-years-magnitude exceeds a bit 2~$\mu$emu, or about 7~$\mu$emu at the strongest field.
To be fair, it has to be noted that the data collected in that figure do show pairs or triples of nearly the same reading, indicating a possible kind of a month long  "islands of stability", but that fact can only be known \emph{a posteriori} and any two or even three equal test readings do not preclude a fluctuation occurring an hour later.
The quoted two numbers do not seem to be particularly large, having in mind that they constitute a mere 0.7\% of the total magnetic moment of this test specimen.
But they have to be related to the expected level of $m_{\mathrm{M}}$.
For the sample S$_{\mathrm{M}}$ the bare magnitudes of $m_{\mathrm{M}}(H)$  are presented in Fig.~6~(a) in ref.~\cite{Gas:2019_MST}.
It reads that $m_{\mathrm{M}}$ of this sample saturates  at  $T = 5$~K and $H=70$~kOe at about 5~$\mu$emu, a value even twice smaller than the magnitude of a possible non-diamagnetic contribution of the bulk sapphire substrate and comparable to the magnitude of the temporal fluctuations of the magnetometer unit.
A nearly identical saturation level is found for two independent measurement performed for two orthogonal orientations of the sample with respect to $H$  (results of analogous measurements taken at 2~K, after further recalculation, are presented here in Fig.\ref{Fig:M(H)} and discussed later).
It should be noted that it usually takes up to 2 days to acquire magnetization curves at the required few temperatures for one orientation of the specimen.
So the instabilities are likely to occur in the experimental time frame of the whole project and it becomes obvious that any such "tiny" instability of the magnetometer will have a destructive effect if such a study is performed using the traditional approach.

In contrast, when the bulk of the substrate is \emph{in situ} compensated in CSH by, say 90\%, then the detrimental contribution of even the maximum magnetometer instability is reduced tenfold, i.e.~its magnitude is reduced to 0.5~$\mu$emu at 70~kOe or only to 10\% of $m_{\mathrm{M}}$.
Certainly, the improvement in reliability of the determination of $m_{\mathrm{M}}$ increases with the degree of compensation.
Surely, we neglect here a marginal influence of the magnetometer instability on the magnitude of $m_{\mathrm{M}}$, this is anyway well below the experimental noise level for the materials studied by this method.
Interestingly, the method lowers the experimental noise below 0.5~$\mu$emu at 70~kOe, approximately 2 to 5 times in regard to the standard method.
This bonus effect stems from the fact that the magnetometer unit processes 10 to 20 times smaller signals, so the moment-establishing fitting procedure works with data sets ("scans") devoid of  large scale fluctuation brought about by the large magnitude signal from the substrate.
Surely, this beneficial effect is limited by the background noise of the magnetometer.

The approach presented here  has already proved particularly useful in investigations of other systems exhibiting minute magnetic signals, i.e. magnetic nanocrystals \cite{NavarroQ:2019_PRB,Mazur:2019_PRB,Navarro-Quezada:2020_Materials}, and very thin antiferromagnetic layers \cite{Wang:2020_PRB,Scheffler:2020_PRM}.
Its usefulness is in practice inversely proportional to the ratio of the magnetic signal of interest to the signal of the substrate at given magnetic field.
Importantly, the whole discussion and the compensational approach presented above applies in the same extent to other popular semiconductor substrates, GaAs, Si, InAs, and others.

The magnetic measurements are performed using a Quantum Design MPMS XL-7 SQUID magnetometer between 2 and 300~K and up to 70~kOe.
For the studies the material is diced into $5 \times 5$~mm$^2$ specimens, which for detailed orientation-dependent studies are further cut into four $\sim$$1.2 \times 5$~mm$^2$ strips.
The experimental material is extracted from the center of the wafers to minimize the inhomogeneity of the Mn doping caused by an inevitable temperature inhomogeneity over the substrate surface during the MBE growth \cite{Gas:2018_JALCOM,Gas:2020_JALCOM}.
Before the measurements, the pieces are bathed in HCl for about 30~mins in an ultrasonic cleaner to remove any metallic contaminants.
The strips are then glued together to form a cuboid of a square cross-section ($1.2 \times 1.2\times 5)$~mm$^3$, as detailed in ref.~\cite{Sztenkiel:2016_NC}.
We use a strongly diluted GE varnish for all gluing purposes \cite{GE_varnish}.
Magnetic studies in both, in-plane and perpendicular orientation are performed by a 90$^o$ rotation of such a cuboidal sample along its length.
This approach assures the same magnitude of the sample to SQUID-pick-up-coil coupling factor $\gamma$.
By definition $\gamma=1$ for a point object.


The importance of this size dependent factor seems to be largely disregarded by the community, whereas it does play a crucial role in precision integral magnetometry.
For a typical $5 \times 5$~mm$^2$ sized sample $\gamma$ differs between the in-plane and the perpendicular orientations by as much as $\Delta \gamma \simeq 0.05$ \cite{Stamenov:2006_RSI,Sawicki:2011_SST}, meaning that for any material the SQUID output in the greater case will be about 5\% larger than in the former.
The scale of the impact of $\gamma$ on the results of MA studies depends on the ratio of $m_{\mathrm{M}}$ to the product of $\Delta \gamma$ and the magnitude of the moment exerted by the substrate at the considered $H$.
For the numbers relevant here the possible change of the signal specific to the substrate alone at $H=70$~kOe due to such a change of $\gamma$ exceeds about 5 times the expected here magnitude of $m_{\mathrm{M}}$.
Therefore, transforming the flat specimen to the cuboidal shape \cite{Sztenkiel:2016_NC}, for which $\Delta \gamma = 0$, eradicates this detrimental contribution, rendering the orientation-dependent measurements in CHS possible at the same time.


The near zero-magnetic-field conditions, which are required to determine the magnitude of the Curie temperature $T_{\mathrm{C}}$, are established by a careful degaussing of the superconducting magnet of the magnetometer.
This is done by applying a slowly oscillating magnetic field with decreasing amplitude and is followed by a soft quench.
As established separately by measuring a pellet of a strong paramagnetic salt (Dy$_2$O$_3$) the typical magnitude of $H$ in the sample chamber does not exceed 0.12~Oe after this procedure.
As reasoned before, all the measurements are performed using the dedicated compensational sample holder depicted in Fig.~\ref{Fig:M(H)}~(a).

An example of the development of the magnetic response of the heretostructures studied here is given for sample S$_{\mathrm{M}}$ in Fig.~\ref{Fig:M(H)}.
The magnetic field dependence of the magnetization $M$, $M(H)$, in the wide field range at 2, 5, 50 and 300 K is presented in panel (b).
It is clearly seen that at the two highest temperatures the signal is paramagnetic, showing no tendency towards saturation in the whole available range of $H$.
Most importantly, no FM-indicative features, like a non-zero remnant magnetization or nonlinearities of $M(H)$ are observed there.
This is the common behavior found in all the studied samples.
It confirms that neither ferromagnetic Mn-enriched clusters are present in these structures nor a carrier induced room temperature ferromagnetism has been induced.
The intimate proximity of the (Ga,Mn)N to the Mg-doped cladding layers has not resulted in any qualitative change of the magnetic character of the (Ga,Mn)N material researched here.

Conversely, at the low-$T$ end, in agreement to the previous studies \cite{Sarigiannidou:2006_PRB,Sawicki:2012_PRB,Kunert:2012_APL,Stefanowicz:2013_PRB,Gas:2018_JALCOM,Kalbarczyk:2019_JALCOM,Sztenkiel:2020_NJP}, $M(H)$ acquires a clear sigmoidal shape and the magnetic saturation is evidenced above 40~kOe at 5~K and around 20~kOe at 2~K.
This behavior indicates that a FM scenario does take place here, but only at lowest temperatures.
A vital clue about the possible nature of this coupling comes from the magnetic anisotropy data presented in Fig.~\ref{Fig:M(H)}~(c).
The most important feature in this graph is the easy-plane character of MA.
This behavior decisively rules out a statistically significant presence of Mn ions in +2 oxidation state, as this orbital singlet state is isotropic and its spin couples antiferromagnetically with the neighbors \cite{Zajac:2001_APL79,Granville:2010_PRB,Bonanni:2011_PRB}.
Therefore,  on the account of the accumulated body of evidences in the low to medium doped (Ga,Mn)N, i.e. for $x$ not greater than several \% \cite{Bonanni:2011_PRB,Stefanowicz:2013_PRB,Gas:2018_JALCOM,Kalbarczyk:2019_JALCOM,Sztenkiel:2020_NJP}, the combined data presented in Fig.~\ref{Fig:M(H)} constitute the telltale evidence of the prevalence of the Mn$^{3+}$ state of Mn ions in the (Ga,Mn)N layer. In such a case the magnetic superexchange is the driving mechanism for the FM witnessed here \cite{Anderson:1950_PR,Goodenough:1958_JPCS,Kanamori:1959_JPCS,Blinowski:1996_PRB}.

Having established the dominant Mn oxidation state in our layers, the Mn concentration can be determined using the crystal field model (CFM) relevant for Mn$^{3+}$ \cite{Gosk:2005_PRB,Mac:1994_PRB,Twardowski:1993_JAP,Herbich:1998_PRB,Wolos:2004_PRB_70,Savoyant:2009_PRB,Stefanowicz:2010_PRBa,Bonanni:2011_PRB,Rudowicz:2019_JMMM}.
The direct CFM computations for noninteracting Mn$^{3+}$ cations yield that the near-saturation value at 70~kOe and 2~K is 3.78~$\mu_B$ per Mn ion \cite{Gosk:2005_PRB,Stefanowicz:2010_PRBa,Sztenkiel:2020_NJP}.
The recently elaborated extension of this approach, which includes the near-neighbor coupling up to the clusters of four Mn$^{3+}$ ions, does not affect the saturation level substantially \cite{Sztenkiel:2020_NJP}.
Taking into account the details of the statistical distribution specific for different-size groups of Mn ions in a 5\% wurtzite crystal \cite{Shapira:2002_JAP}, one finds that the saturation level increases only to 3.79~$\mu_B$.
Acknowledging, that in fact the main error in the determination of saturation magnetization originates from a large, about 20\%, uncertainty of the (Ga,Mn)N layer thickness, the single ion figure has been adopted, as it is $x$-independent.

\begin{figure}[htb]
\centering
\includegraphics[width=9 cm]{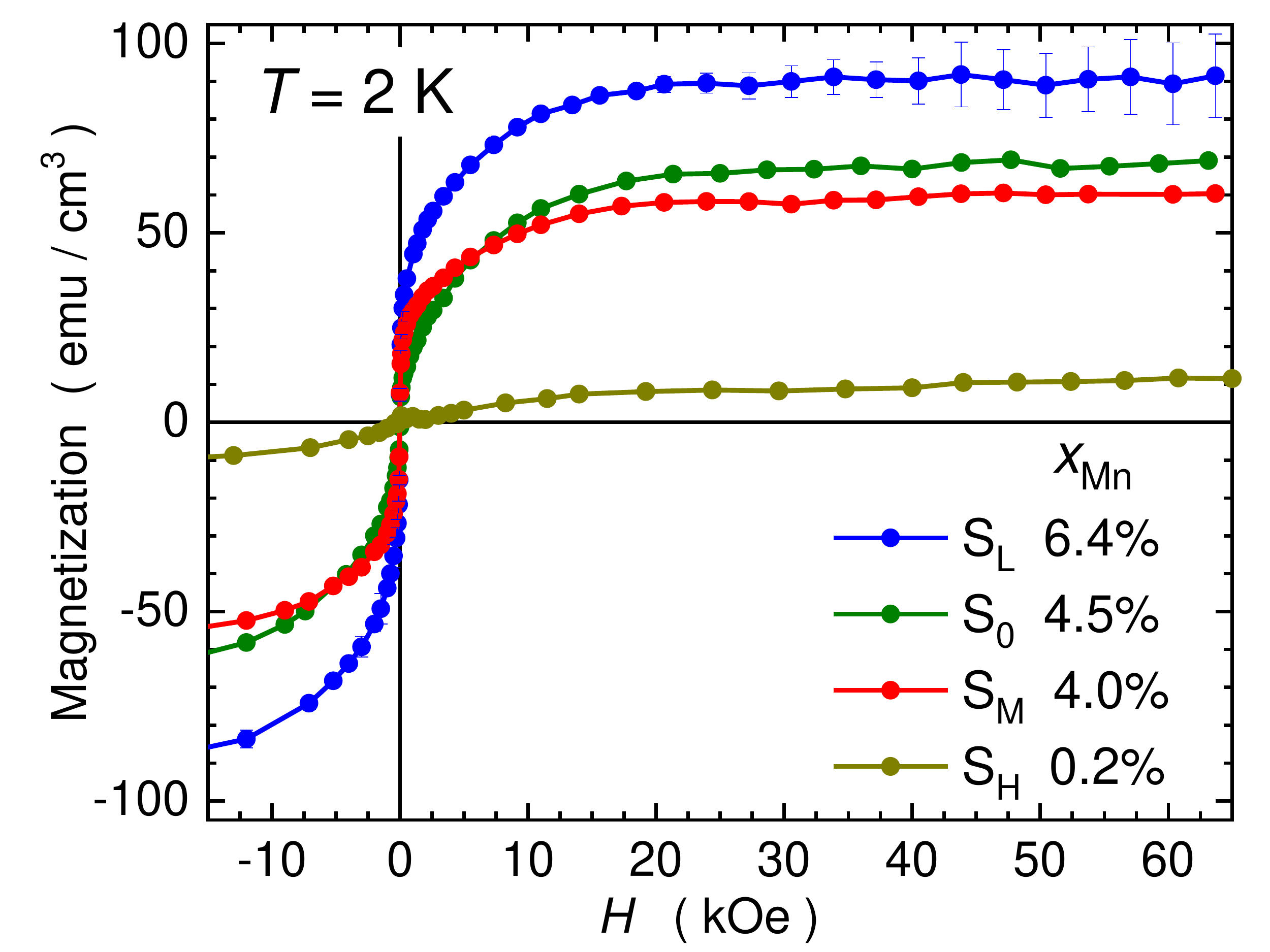}
\caption{\label{Fig:M(H,x)} (Color on line) Isothermal in-plane magnetization curves for all investigated structures measured at 2~K. The established saturation levels are used to calculate Mn concentration assuming nominal layer thickness of 4.3~nm and Mn$^{3+}$ state of Mn ions, i.e.~exibiting saturation at 3.79~$\mu_B$/Mn \cite{Sztenkiel:2020_NJP}. For clarity experimental errors are indicated only for one sample.}
\end{figure}
$M(H)$ curves collected at 2~K for all the samples  are given in Fig.~\ref{Fig:M(H,x)}, from where the saturation levels can be read and magnitude of $x$ established.
These are listed in the legend and in Table~\ref{tab:Samples}.
While the magnitude of $x$ for three of the layers stays at similar level between 4 and 6.4\%, the low value established for S$_{\mathrm{H}}$ layer indicates that the Mg flux of $2\times 10^{-8}$~Torr BEP appears to be too high to incorporate Mn in the successive layer in an appreciable quantity.
It is understood, following the previous finding reported by Tropf \emph{et al.} \cite{Tropf:2016_JCG}, that a roughening of the surface expected on approaching the conditions specific to the GaN surface polarity inversion made the conditions for Mn incorporation less favorable.
Indeed, Mn incorporation in GaN decreased by 100 to 1000 times after the transition to N-polar case induced by an excessive Mg flux  \cite{Tropf:2016_JCG}.

\begin{figure}[htb]
\centering
\includegraphics[width=13.5 cm]{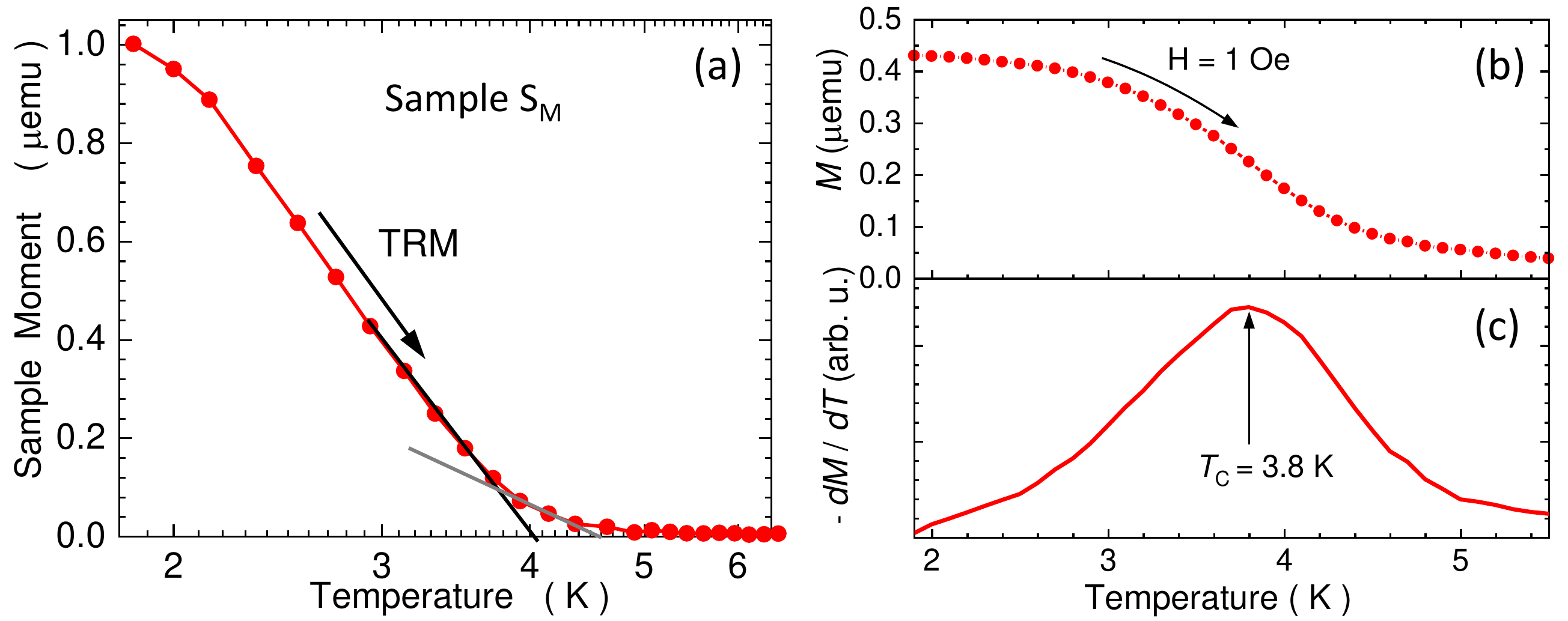}
\caption{\label{Fig:TRM} (Color on line) Critical behavior of sample S$_{\mathrm{M}}$ and the Curie point determination from (a) thermoremanent magnetization (TRM) and (b) from the position of the inflection point  on the temperature dependence of magnetization $M(T)$. In panel (a) the two lines indicate two possible methods of the interpolation of TRM to zero. They yield the lower and the upper bound limits of $T_{\mathrm{C}}$ from this method. The magnitude of the inflection temperature is obtained from the position of a maximum on numerically computed $-dM(T)/dT$, presented in panel (c). The $M(T)$ dependency is measured at $H = 1$~Oe.}
\end{figure}
The Curie temperature of the (Ga,Mn)N layers is determined by two different methods employed previously \cite{Stefanowicz:2013_PRB,Gas:2018_JALCOM}.
In either case, before the measurement, the superconducting magnet of the magnetometer is degaussed following the recipe given previously.
In the first approach, the sample is cooled down from room temperature to the base temperature of $T = 1.8$~K in $H = 1$~kOe.
Then, the magnetic field is quenched using the soft quench of the superconducting magnet.
Under these conditions, the thermoremanent moment (TRM) is collected on increasing temperature until it drops to zero (that is, decidedly below the noise level).
This temperature is understood to represent the $T_{\mathrm{C}}$ of the system.
An example of the TRM trace for sample S$_{\mathrm{M}}$ is depicted in Fig.~\ref{Fig:TRM}~(a).
A clear signal is seen at the base temperature, whereas at the transition region (between 4 and 4.8 K) the disappearance of TRM is not particularly abrupt.
This smearing out of the transition is a characteristic feature of dilute ferromagnets with randomly distributed magnetic species coupled in a percolation fashion via a short-range coupling.
The formation of the FM state below the percolation limit (about 18\%) is facilitated in only a few per cent (Ga,Mn)N by (i) exclusively positive values of the spin-spin coupling constants (at least up to the 16th neighbor \cite{Simserides:2014_EPJ-WoC}), and by (ii) a power-law decay of the exchange constant with distance.
The process of the extrapolation TRM to zero can be aided by plotting the TRM data in the $\log(T)$ scale.
From the two "free hand" interpolation examples presented in Fig.~\ref{Fig:TRM}~(a) we obtain $4<T_{\mathrm{C}}<4.6$~K.

In the second method, exemplified in Fig.~\ref{Fig:TRM}~(b) $T_{\mathrm{C}}$ is determined from the inflection point of the $M(T)$ dependency.
It is measured at a very low field, $H = 1$~Oe, and the inflection temperature is determined from the position of a maximum of electronically computed $-dM/dT$, as shown in Fig.~\ref{Fig:TRM}~(c).
This method yields $T_{\mathrm{C}} = 3.8 \pm 0.3$~K.
For further quantitative analysis of the magnetic results we take the average of these two magnitudes  with an error bar reflecting the maximum spread of the possible $T_{\mathrm{C}}$ values.
In the case of the S$_{\mathrm{M}}$ sample $T_{\mathrm{C}} = 4 \pm 0.5$~K.
All the $T_{\mathrm{C}}$ values are listed in Table~\ref{tab:Samples}.

\begin{figure}[htb]
\centering
\includegraphics[width=7cm]{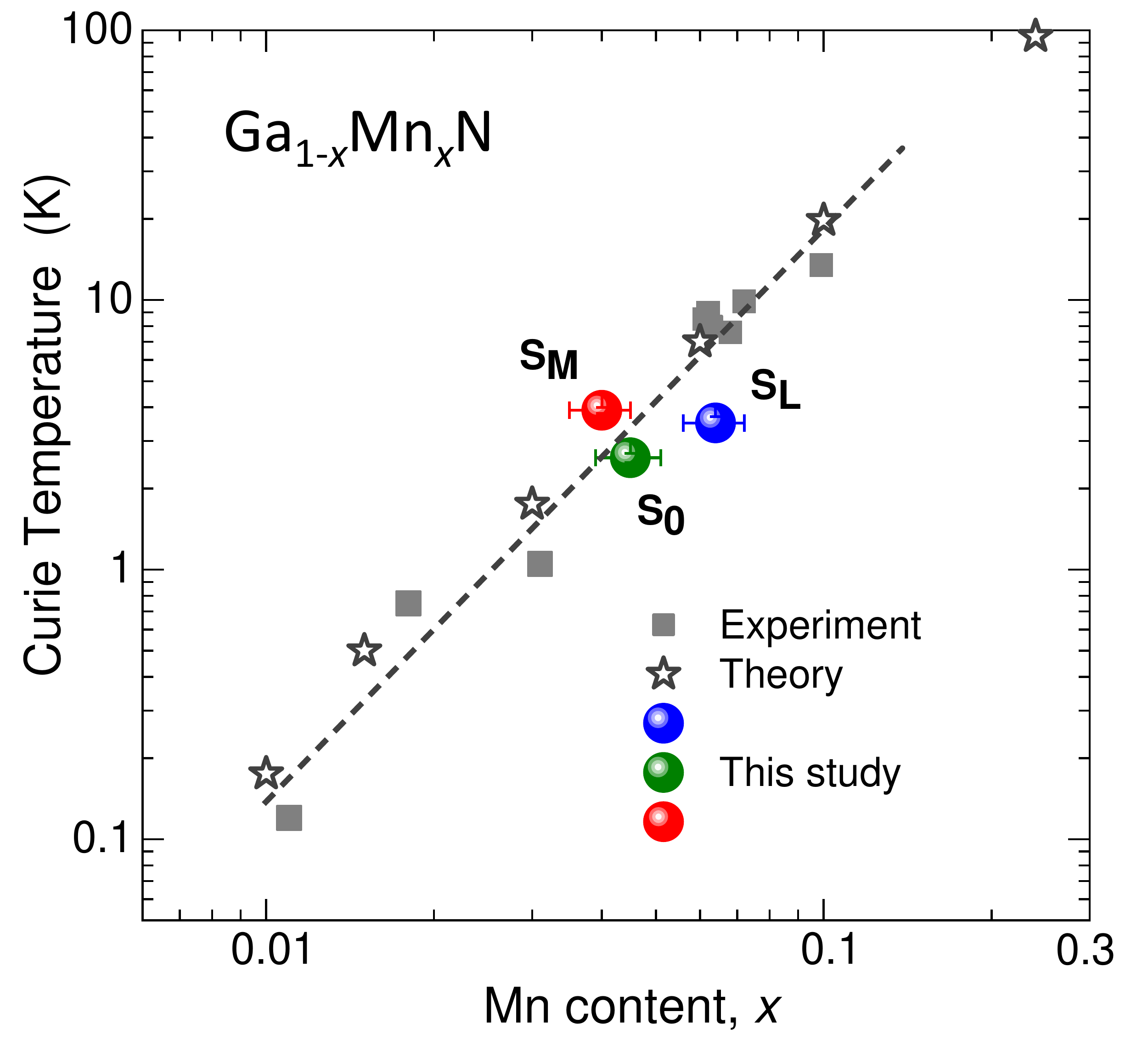}
\caption{\label{Fig:Ph_diag} (Color on line) Experimentally found Curie temperatures of the layers studies here (bullets) located on the background of Ga$_{1-x}$Mn$_x$N magnetic phase diagram (gray symbols) \cite{Stefanowicz:2013_PRB}. Full squares represent experimental points, stars - results of Monte Carlo simulations with exchange integrals from the tight-binding model. The dashed line is a guide to the eye.}
\end{figure}
The magnetic characteristics of the (Ga,Mn)N layer determined in this study are summarized in the magnetic phase diagram presented in Fig.~\ref{Fig:Ph_diag}.
The background data (gray symbols) are taken from ref.~\cite{Stefanowicz:2013_PRB}.
Only three samples with $T_{\mathrm{C}}$ lying in the experimentally available $T$-range, $T \geq 2$~K, are presented there.
We find that these layers group very close to the trend established for thick single (Ga,Mn)N layers, fabricated just on GaN-buffered sapphire.
This finding indicates that our attempt to augment the FM spin-spin coupling by placing the (Ga,Mn)N layer with an intimate contact with Mg-doped GaN has not resulted in any change of the  already known  magnetic properties of this material.
The most likely reason for this unaltered magnetic picture found here is attributed to the presence of residual Mn in the Mg-doped layers.
This is evidenced in Figs.~\ref{Fig:TEM}~and~\ref{Fig:SIMS}.
Acting as deep traps, the Mn species deactivate hole states formed by the Mg-doping.
Indeed, the doping with Mn has become one of the most promising methods to produce industry standard semi-insulating bulk GaN substrates for high power and high frequency electronics \cite{Zajac:2018_PCGCM,Bockowski:2018_JCG}.
Mn doping levels as low as $2 \times 10^{17}$~cm$^{-3}$ has been proved sufficient to produce highly resistive GaN even at room temperature \cite{Iwinska:2019_JJAP}.

\section{Conclusions and Outlook}

Specially designed GaN:Mg~/~(Ga,Mn)N~/~GaN:Mg structures on GaN-buffered Al$_2$O$_3$ substrates have been grown by molecular beam epitaxy following state-of-the-art growth protocols, aiming at an altering of the magnetic spin-spin interaction in the (Ga,Mn)N layer by the indirect co-doping  by holes from the cladding layers.
The four structures prepared for this study have been carefully structurally characterized.
In particular none of them exhibited any foreign Mn-rich phases.
To eliminate spurious sources of the magnetic signals from the substrates,  a dedicated experimental approach has been applied increasing both, the credibility and the sensitivity of the integral magnetometry at strong magnetic field.
The magnetic characteristics proved identical to the already known properties of the thick (Ga,Mn)N single layers, the superexchange driven low temperature ferromagnetism at low end of cryogenic temperatures and purely paramagnetic response above.
The possible cause of a lack of any effects brought about by the adjacent Mg-doping is the presence of residual Mn in the cladding layers, resulting in the deactivation of the p-type doping intended there.
However, the exact cause of this has not been identified, the (out-)diffusion of Mn from the central (Ga,Mn)N layer has been evaluated to be insufficient at the range of the growth temperatures specific for the MBE technique.
This finding points out on the yet another  technological challenge lying ahead of such, or similar, attempts aiming at inducing a carrier mediated ferromagnetic coupling in GaN, especially at elevated temperatures.

However, it has to be underlined that the non optimistic results reported here do not preclude a carrier mediated ferromagnetism, or more generally speaking, a homogeneous ferromagnetism in magnetic-element-doped GaN at technology relevant high temperatures.
On the contrary, taking on Mn as an representative example, the following avenues (ordered by the ascending degree of speculativeness) could be exercised with a fair chance of success.
(i) The use of higher quality substrates. In particular GaN homoepitaxy. It may be expected that a higher structural quality material (with some orders of magnitude smaller density of dislocations) might provide less disturbance of magnetic coupling between Mn ions in proximity, leading to a stronger coupling.
(ii) Active Fermi level engineering during the growth. There are at least two possibilities here. (iia) By co-doping with hydrogen, which is a problem for MBE, but can be solved; or (iib) by irradiating the growth front with UV light. Such a photo-assisted growth increases the incorporation of less soluble elements. The resulting increased position of the Fermi level is expected to lead to a higher amount of incorporated Mn ions in (Ga,Mn)N. In either case a control over the Fermi level can influence the lattice position (e.g., interstitial versus substitutional) and the distribution of magnetic impurities over cation sites \cite{Dietl:2006_NM}. Both solutions require substantial modifications of the existing growing facility, but can be realized.
(iii) Ion implantation. This technique is known to produce virtually any combination of material-dopant(s) system, proved particularly effective in the case of (Ga,Mn)P \cite{Scarpulla:2005_PRL}, but also with arsenide DMS \cite{Scarpulla:2003_APL,Zhou:2009_PRB,Yuan:2017_PRM}. The main hurdle to overcome here is the restoration of the high crystalline quality of the severely ion-damaged host, and nanosecond short pulse laser annealing seems mandatory to achieve a high quality homogeneous material \cite{Yuan:2016_AP&I}. Otherwise, during far longer rapid thermal processing the implanted atoms tend to diffuse out or to agglomerate into precipitations within the material \cite{Pereira:2013_JAP}. Encouragingly, a homogeneous p-type hyperdoping has been shown possible in Ga-implanted Ge \cite{Prucnal:2019_PRM}.
(iv) Additional weak co-doping with  As (about $0.1-1$\%). Arsenic is known to cause a considerable up-shift of the valence band \cite{Grodzicki:2019_V,Kudrawiec:2020_APR}. So, its presence may also provide a strong effect on the electronic properties of the Mn ions, possibly even promoting holes within the Mn-band. The main hurdle to overcome here will be the detrimental impact of the incorporation of As on the material quality.
(v) The same As can be used as the surfactant to initiate growth of cubic (Ga,Mn)N on GaAs substrates \cite{Novikov:2005_JVST-B}. Such a form of (Ga,Mn)N has been shown to exhibit p-type conductivity above 150~K \cite{Edmonds:2005_APL} and ferromagnetism at single Kelvin temperature range \cite{Sawicki:2004_ICPS}.
(vi) Strain engineering. It was postulated that inducing a tensile strain along the wurtzite c axis might lead to sufficient delocalization of hole states and so long-range hole-mediated magnetic interactions \cite{Raebiger:2018_PRM}.
(vii) The last on this list, but by no means the least avenue would be a fine tuning of the donor co-doping to half-empty the Mn mid-gap level so the concentrations of Mn$^{3+}$ and Mn$^{2+}$  ions are approximately equal.
Such conditions favor double exchange mechanism in which much higher coupling temperatures are envisaged than for the short range superexchange \cite{Krstajic:2004_PRB}.
However, the net gain in the FM interaction strength can be diminished by the antiferromagnetic coupling between Mn$^{2+}$ pairs.

Certainly, any combination of the approaches listed above may serve better than a single one of them.
This would depend on the available resources.
In any case, to validate the results, the research has to include an adequate and dedicated nano-characterization and the magnetic studies have to be performed with a great attention to the details.

\section*{Declaration of competing interest}

The authors declare that they have no known competing financial interests or personal relationships that could have appeared to influence the work reported in this paper.

\section*{Acknowledgments}

All authors contributed to the preparation of the draft of the manuscript and read, commented and corrected its final version.
The work has been supported by the National Science Centre (Poland) through project OPUS 2018/31/B/ST3/03438.


%

\end{document}